**Moiré-Induced Magnetoelectricity in Twisted Bilayer NiI$_2$**

Haiyan Zhu[a], Hongyu Yu[a], Weiqin Zhu[a], Guoliang Yu[a], Changsong Xu[a,b,*], Hongjun Xiang[a,†]

[a]*Key Laboratory of Computational Physical Sciences (Ministry of Education), Institute of Computational Physical Sciences, State Key Laboratory of Surface Physics, and Department of Physics, Fudan University, Shanghai 200433, China*

[b]*Hefei National Laboratory, Hefei 230088, China.*

[*]Email: csxu@fudan.edu.cn

[†]Email: hxiang@fudan.edu.cn

**Abstract**

Twisted magnetic van der Waals (vdW) materials offer a promising route for multiferroic engineering, yet modeling large-scale moiré superlattices remains challenging. Leveraging a newly developed SpinGNN++ framework that effectively handles spin-lattice coupled systems, we develop a comprehensive interatomic machine learning (ML) potential and apply it to twisted bilayer NiI$_2$ (TBN). Structural relaxation introduces moiré-periodic "bumps" that modulate the interlayer spacing by about 0.55 Å and in-plane ionic shifts up to 0.48 Å. Concurrently, our ML potential, which faithfully captures all key spin interactions, produces reliable magnetic configurations; combined with the more accurate generalized KNB mechanism, it delivers precise spin-driven polarization. For twist angles $1.89° \leq \theta \leq 2.45°$, both mechanisms become prominent, yielding rich polarization textures that combine ionic out-of-plane dipoles with purely electronic in-plane domains. In the rigid (unrelaxed) bilayer, skyrmions are absent; lattice relaxation is thus essential for generating polar-magnetic topologies. In contrast, near $\theta \approx 60°$, stacking-dependent ferroelectric displacements dominate, giving rise to polar meron-antimeron networks. These results reveal cooperative ionic and spin-driven ferroelectricity in TBN, positioning twisted vdW magnets as adaptable platforms for tunable multiferroic devices.

**Introduction**

Two-dimensional (2D) van der Waals (vdW) materials provide a rich platform for engineering exotic quantum phases, including long-range ferromagnetic (FM)[1, 2] and ferroelectric (FE)[3-5] orders. However, realizing magnetoelectric coupling in such systems remains a significant challenge. Multiferroics, known for their remarkable static[6, 7] and dynamic[8, 9] magnetoelectric properties, can offer promising solutions. In type-II multiferroics, the coupling between magnetic and dipolar order parameters—mediated through spin-current mechanisms or the inverse Dzyaloshinskii–Moriya interaction—results in strong magnetoelectric effects[10, 11]. In particular, a unified polarization model[12] has been proposed, in which the general spin-current mechanism systematically accounts for the magnetically induced ferroelectricity observed in multiferroics such as $CuFeO_2$[13] and $NiI_2$[14, 15]. Among these materials, $NiI_2$ is the first well-documented 2D type-II multiferroic, where helical magnetic ordering and strong spin-orbit coupling combine to generate substantial magnetoelectric coupling[11, 14, 16-21]. $NiI_2$ adopts a $CdI_2$-type layered structure and a triangular arrangement of $Ni^{2+}$ ions in edge-sharing $NiI_6$ octahedra. Previous work[20] has developed a realistic spin Hamiltonian for bulk $NiI_2$, accurately capturing the essential features of its experimental helical ground state. Recent studies[14, 19] have further demonstrated that the helimagnetic state persists down to the monolayer limit below $T_N \sim 20$ K, with potential for tuning multiferroic order *via* strain[22], pressure[23, 24], and substrate engineering[14].

In addition to these approaches, stacking order has proven to be an effective strategy for controlling FM and FE properties in vdW materials[3, 25, 26]. Specifically, twisted vdW systems, where moiré superlattices emerge due to a small twist angle between layers, create a diverse array of stacking domains and domain walls that substantially affect magnetic and polar properties[27-31]. For instance, magnetic moiré domains have been experimentally observed in twisted bilayer $CrI_3$ at small twist angles[31-33]. Recent theoretical efforts[34] have predicted skyrmion phases and out-of-plane polarization in twisted bilayer $NiI_2$ (TBN) using simplified Hamiltonian models. However, previous models for twisted systems[30, 34, 35] often neglect important effects, such as intrinsic ferroelectric displacements and spin-lattice coupling, all of which become critical at small twist angles. As a result, little is understood about how moiré-scale structural relaxation affects magnetic properties in TBN, especially at small twist angles, or how polarization emerges from the system's multiferroic nature.

In addition to the polarization arising from spin-driven ferroelectricity, non-polar monolayers can develop in-plane polarization (IPP) or out-of-plane polarization (OPP) once stacking arrangements break inversion symmetry[3, 4, 36, 37]. For example, in bilayer hexagonal boron nitride

(h-BN)[3] and transition metal dichalcogenides (TMDs)[4, 36, 37], the magnitude and direction of polarization are sensitive to relative sliding between layers. Unlike conventional ferroelectrics with discrete polar states, 2D systems form spontaneous polarization domains driven by dimensional constraints, elastic strain, and interfacial charge effects[38, 39], giving rise to topological structures like skyrmions[40], merons[41], and vortices[39]. In $NiI_2$ monolayers, which possess inversion symmetry (point group $D_{3d}$), IPP, OPP, or combined polarization (CP) with both in-plane and out-of-plane components can appear in bilayers when anti-aligned with a 60° rotation relative to the bulk configuration[42]. Therefore, strategically twisting such bilayers by nearly 60° can harness local symmetry breaking to create moiré polar domains. These topological polar textures in TBN present intriguing possibilities for experimental detection, manipulation, and device applications[39, 41, 43], warranting further investigation.

In this Letter, we apply the time-reversal E(3)-equivariant neural network and SpinGNN++ framework [44] to investigate the large-scale moiré system of TBN across commensurate twist angles ranging from 1.09° to 21.79° and 38.21° to 58.91°. The resulting spin-lattice potential effectively captures both the structural and magnetic properties of TBN, including moiré-periodic "bumps" ($\approx$ 0.55 Å) and in-plane ionic shifts up to 0.48 Å. We demonstrate that structural relaxation induces symmetry breaking and generates OPP in TBN. At twist angles between 1.89° and 2.45°, spin spirals locked to moiré domains lead to new polarization patches originating from the electronic contribution. While for TBN~60°, inversion symmetry breaking due to anti-aligned stacking creates topological meron-antimeron networks. These results highlight the interplay of ionic and spin-driven ferroelectricity in twisted vdW magnets.

**Results**

**Twist-free bilayer $NiI_2$**

We first establish the structural and magnetic properties of twist-free bilayer $NiI_2$ using density functional theory (DFT) calculations [see Supplemental Material (SM)[45] for details]. Considering the intrinsic symmetry of monolayer $NiI_2$, we identify three high-symmetry stacking configurations: AA, AB, and AB' (all with space group $P\bar{3}m1$), each corresponding to different relative shifts of the top layer with respect to the bottom layer. These shifts are (0,0), (1/3, 2/3), and (2/3, 1/3), respectively [see left panels of FIG.1(a)]. When the top layer is rotated by 60° (anti-aligned) relative to the bottom layer, analogous shifts produce the R-AA ($P\bar{6}m2$), R-AB ($P3m1$), and R-AB' ($P3m1$) configurations [see right panels of FIG. 1(a)]. The AB stacking corresponds to the natural stacking in bulk NiI$_2$. The optimized lattice constant $a = b =$ 3.97 Å agrees well with the experimental value of $a = b =$ 3.91 Å[14].

To explore the energetics, we perform rigid shifts of one layer along the $a$ and $b$ axes and calculate the energy relative to the AB (or R-AB) reference configuration. The energy differences are shown in FIG. 1(b). Each structure is modeled with intralayer FM and interlayer antiferromagnetic (AFM) order for simplicity. The results show that AA and AB (and likewise R-AB and R-AB') stackings are low-energy stackings, while AB' (R-AA) stacking is less stable, lying 9.68 (9.87) meV/atom higher in energy compared to AB (R-AB). Across all stacking shifts, AFM interlayer coupling is preferred [FIG.1(c)]. The interlayer distance follows a similar trend to the energy landscape, with unfavorable stackings exhibiting larger spacing between layers [FIG.1(d)].

The spin Hamiltonian of AB-stacked bilayer NiI$_2$ is obtained using DFT and four-state energy mapping method[46]:

$$H = \sum_{\langle i,j \rangle_1}[\boldsymbol{S_i} \cdot \mathbb{J}_1 \cdot \boldsymbol{S_j} + B(\boldsymbol{S_i} \cdot \boldsymbol{S_j})^2] + \sum_{\langle i,j \rangle_3} J_3 \, \boldsymbol{S_i} \cdot \boldsymbol{S_j} + \sum_{\langle i,j \rangle_{1,2}} J_{nl} \, \boldsymbol{S_i} \cdot \boldsymbol{S_j} + \sum_i A_z \, S_i^z S_i^z \quad (1)$$

where $\mathbb{J}_1$ is a $3 \times 3$ matrix representing the full second order exchange interactions for the first nearest neighbor. Key parameters include isotropic Heisenberg exchange $J_1 = -3.547$ meV, the Kitaev interaction $K = 1.441$ meV, a sizable biquadratic term with $B = -0.609$ meV, an antiferromagnetic $J_3 = 2.856$ meV, interlayer couplings $J_{1l} = -0.05$ meV and a strong AFM $J_{2l} = 0.888$ meV, and also the single-ion anisotropy (SIA) $A_z = 0.137$ meV favoring in-plane spin alignment. These values closely match those obtained for bulk NiI$_2$[20].

The magnetic ground state of AB-stacked bilayer NiI$_2$ is determined using Monte Carlo (MC) and conjugate gradient (CG) methods implemented in the PASP software[47] based on the spin Hamiltonian in Eq. (1) (see SM[45]). In agreement with previous works[19, 22], the magnetic ground state of low-dimensional NiI$_2$ is strongly influenced by its geometry, showing nearly degenerate energies for spin spiral propagation along the $\langle 110 \rangle$ and $\langle 1\bar{1}0 \rangle$ directions (FIG. S1[45]). Note that $\langle 110 \rangle$ direction is defined as the nearest Ni-Ni bonding direction in both untwisted and twisted configurations throughout this study. Using an $18 \times 18 \times 1$ supercell, we identify the magnetic ground state of bilayer NiI$_2$ as a proper screw propagating along the $\langle 110 \rangle$ direction with $\boldsymbol{q} \sim (0.222, 0.222, 0)$. This is consistent with the propagation direction reported for monolayer NiI$_2$[19].

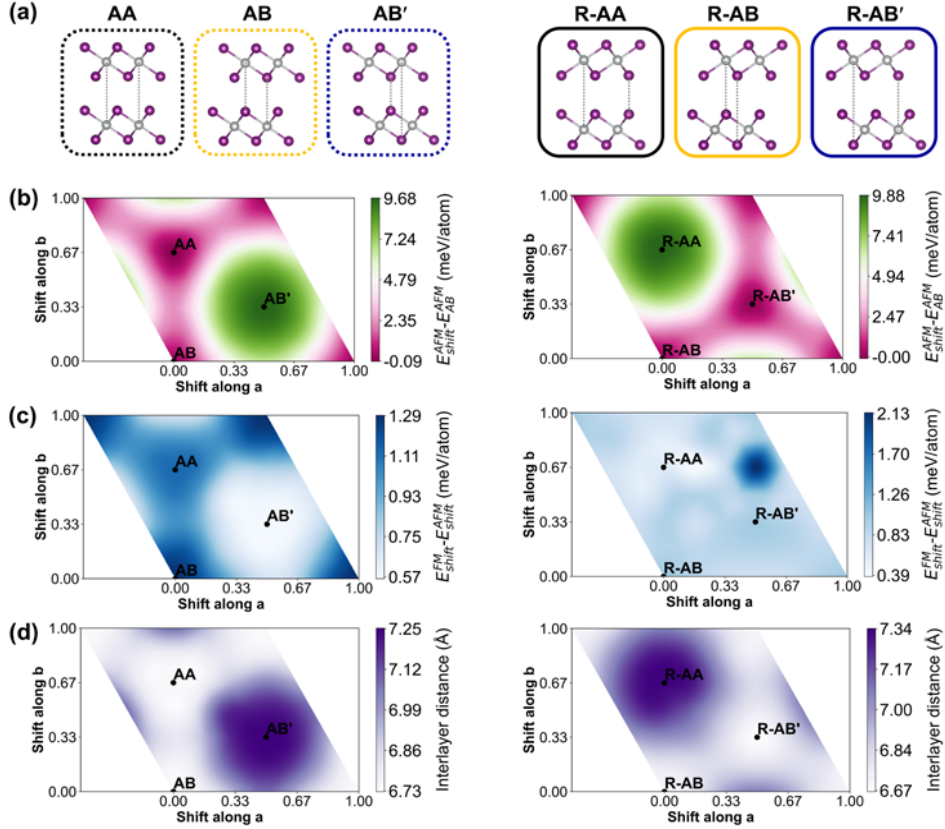

FIG. 1. Stacking-dependent properties of twist-free bilayer NiI$_2$. (a) Schematics of stacking configurations: AA, AB, and AB′, generated by shifting the top layer by (0,0), (1/3, 2/3), and (2/3, 1/3), respectively. The R-AA, R-AB, and R-AB′ configurations (right) arise from a 60° rotation of the top layer and corresponding shifts. (b) Stacking energy as a function of shifting, defined as $\Delta E = E_{stack} - E_{AB}$. Interlayer coupling is fixed to be AFM. (c) Energy differences between FM and AFM interlayer configurations across various stacking orders. (d) Interlayer distances across different stackings, defined as the vertical separation between Ni atoms in adjacent layers.

**Moiré-induced lattice relaxation and its role in ionic polarization in TBN**

The SpinGNN++ magnetic potential allows us to perform large-scale structural relaxations of TBN with near first-principles accuracy at a greatly reduced computational cost. The neural-network potential is trained on 5,981 data points from first-principles bilayer NiI$_2$ calculations, yielding a mean absolute error of 0.076 meV/atom and an $R^2$ score of 0.9999, demonstrating its high accuracy (see SM[45]). This model is used to fully optimize all commensurate TBN structures, covering twist angle $\theta$ from 21.79° to 1.09° (near 0°) and 38.21° to 58.91° (near 60°).

We first exam the symmetry changes in the 2.13° TBN structures. For reference, an isolated NiI$_2$ monolayer has centrosymmetric point group $D_{3d}$, while a rigid 2.13° bilayer already lowers this to the non-centrosymmetric $D_3$. These structures belong to the nonpolar space groups P312. After relaxation, the lattice distortions redistribute the AA, AB and AB′ stacking domains while preserving the out-of-plane $C_3$ axis, thereby enabling an ionic OPP consistent with the generalized stacking ferroelectricity mechanism[42].

To illustrate the effect of lattice relaxation, we compare the relaxed TBN structures at $\theta = 3.89°$ and $\theta = 2.13°$. As shown in FIG.2 (a)(b), the moiré pattern continuously evolves among three primary stacking domains. The low-energy AA and AB regions expand into large triangular domains, while the less favorable AB' areas shrink. In-plane displacements of top-layer Ni atoms in AB′ regions exhibit a helical pattern whose amplitude increases with decreasing $\theta$, reaching 0.48 Å at $\theta = 2.13°$ [FIG.2 (c)(d)]. In contrast, the bottom-layer Ni atoms displace in the opposite sense, following a similar pattern. Interlayer displacements also increase in the AB′ domains, resulting in layer spacing ranging from 6.73 Å to 7.28 Å [FIG. 2(e)(f)], yielding an ionic polarization of $\sim 10 \times 10^{-3} e \cdot Å$ that is comparable to the spin-driven contribution. It eventually results in a net OPP of 0.136 $e \cdot Å$ associated with the $C_3$ symmetry. Moreover, the domain walls connecting AB′ regions become more narrow and sharp at lower twist angles. These results highlight the critical role of moiré-scale lattice relaxation, which has been underexplored due to substantial computational costs, in driving polarization in moiré superlattices.

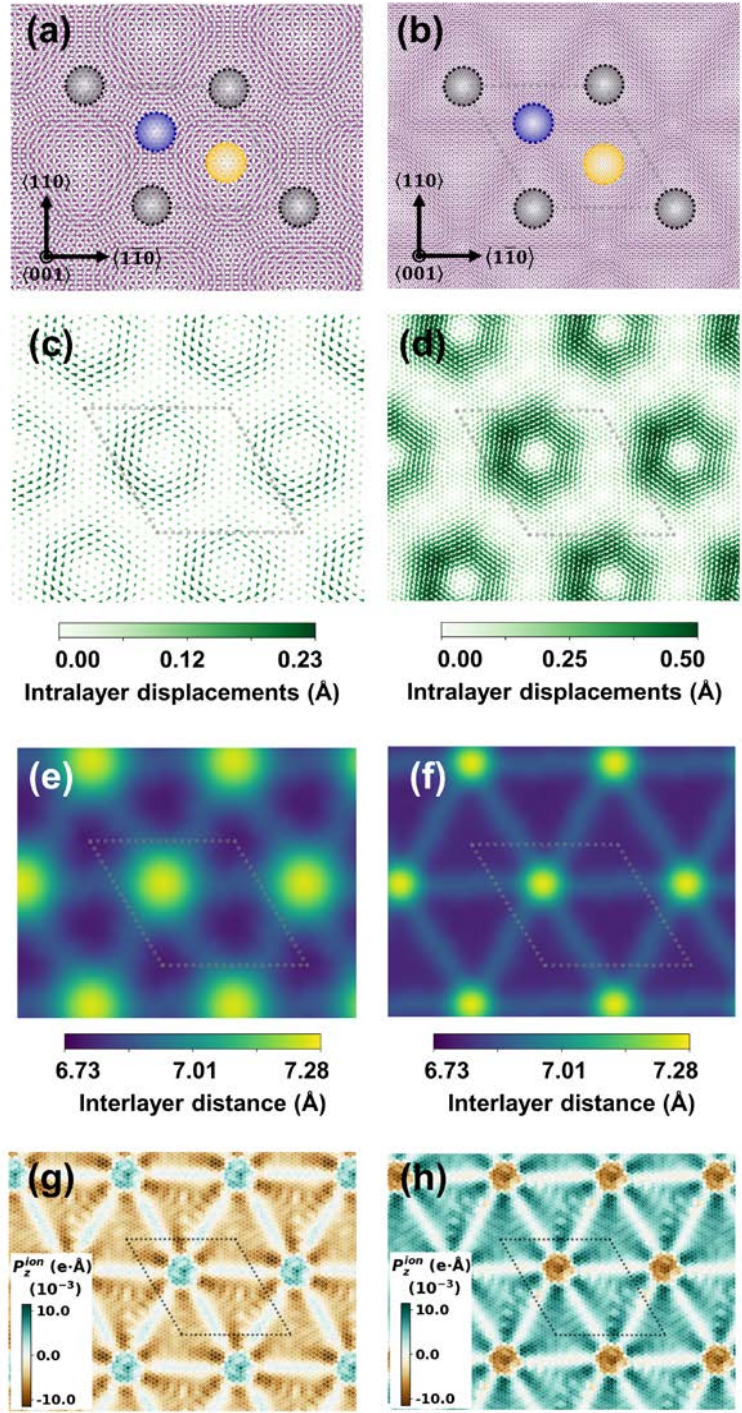

FIG. 2. Structural changes and polarization induced by relaxation in TBN. (a, b) Relaxed TBN structures at $\theta = 3.89°$ and $\theta = 2.13°$, respectively. AA, AB, and AB′ stackings are marked by black, orange, and blue circles (c, d) In-plane displacements of top-layer Ni atoms following relaxation at $\theta = 3.89°$ and $\theta = 2.13°$, respectively. (e, f) Interlayer distances at $\theta = 3.89°$ and $\theta = 2.13°$, respectively. (g, h) OPP patterns in the top and bottom layers at $\theta = 2.13°$, respectively. Gray dashed lines outline the moiré primitive cells.

**Spin and polarization patterns in TBN**

To investigate the magnetoelectric properties of TBN, we first validate our SpinGNN++ potential by comparing its predictions of key magnetic parameters for twist-free bilayer NiI$_2$ with results from DFT. The underlying Hamiltonian explicitly retains intrinsic Kitaev and biquadratic terms together with twist-modulated interlayer exchange, enabling a faithful description of the competition that governs skyrmion stability. Table S1[45] lists the key magnetic parameters for the AB-stacked bilayer extracted from our ML potential. A $20 \times 20 \times 1$ supercell is then employed to optimize both atomic and spin degrees of freedom via annealing and conjugate gradient methods (see SM[45]). The ML potential reproduces the twist-free ground state—a $\langle 110 \rangle$-propagating cycloid with $\mathbf{q} = (0.2, 0.2, 0)$—in close agreement with the DFT and four-state Hamiltonian (Eq. [1]). Similar behavior arises for all aligned stackings, while anti-aligned stackings favor a vertical cycloid along $\langle 110 \rangle$ (VC$^{\langle 110 \rangle}$) with the same period. The strong $J_3/J_1$ ratio indicates robust intralayer frustration across various stackings. Consistent with previous study[19, 20], parameters derived from Perdew-Burke-Ernzerhof (PBE) favor a helical screw along $\langle 110 \rangle$ over a proper screw along $\langle 1\bar{1}0 \rangle$ regardless of whether the four-state or SpinGNN++ approach is used.

Next, we conduct spin-lattice molecular dynamics simulations to examine the evolution of magnetic states across different twist angles in TBN. Each simulation included at least 896 Ni atoms in the moiré superlattice to ensure consistent sampling. As seen in FIG. 3 (and FIG. S7[45]), TBN with $\theta$ >2.45° or $\theta$ <1.89° exhibits spin spirals similar to those in untwisted bilayer NiI$_2$, characterized by in-plane propagation along $\langle 110 \rangle$ with $\lambda \sim 5a$. To quantify the direction and magnitude of the propagation vector $\mathbf{q}$, we calculate the spin structure factor $S(\mathbf{q})$ for each twist angle (FIG. S8[45]). The structure factor is defined as $S(\mathbf{q}) = \frac{1}{N}\Sigma_{\alpha=x,y,z}\langle |\Sigma_i \langle S_i^\alpha \rangle e^{-i\mathbf{q}\cdot r_i}|^2 \rangle$, where $r_i$ is the position of spin $S_i$ and $N$ is the total number of spins in the supercell used for the spin-lattice simulations. The results reveal characteristic peaks corresponding to propagation along the $\langle 110 \rangle$ direction and its equivalents. Due to the near-degenerate energy between IC$^{\langle 110 \rangle}$ and IC$^{\langle 1\bar{1}0 \rangle}$ in NiI$_2$, small domains may form with $\mathbf{q} \parallel \langle 1\bar{1}0 \rangle$ [white boxes, FIG. 3].

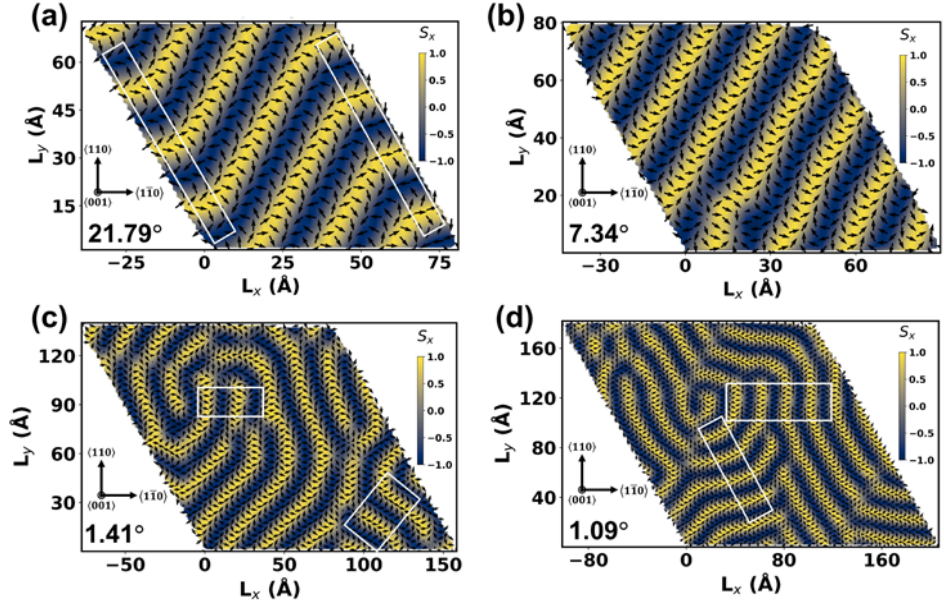

FIG. 3. Spin textures of aligned TBN at various twist angles, obtained from spin-lattice simulations with SpinGNN++. Black arrows represent in-plane spin directions and magnitudes, while the background color maps the $S_x$ component. White boxes highlight regions where the spiral propagation vector aligns with $\langle 1\bar{1}0 \rangle$, in contrast to the dominant $\langle 110 \rangle$ direction.

The range $1.89° \leq \theta \leq 2.45°$ has attracted our attention due to its markedly different behavior. Within this interval, the main propagation direction shifts from $\langle 110 \rangle$ to $\langle 1\bar{1}0 \rangle$ (FIG.S8[45]). Notably, patches with specific $q$-vector directions emerge, forming a spin spiral pattern (SSP) that closely aligns with the moiré potential pattern [FIG.4(a) and FIG. S9[45]]. At $\theta=2.13°$, the pattern achieves its highest order, forming a $C_3$-symmetric domain that appears around the AB' stacking, where six $q$-vector patches converge. Additional $C_3$-symmetric domains are formed near the AA and AB regions [FIG.4(a)]. A controlled calculation at $\theta = 2.13°$ with rigid layers confirms that, in the absence of atomic relaxation, the distinctive spin spiral pattern vanishes and no emergent spin or polarization textures appear (FIG. S10[45]), highlighting the key role of lattice distortions in shaping TBN magnetism.

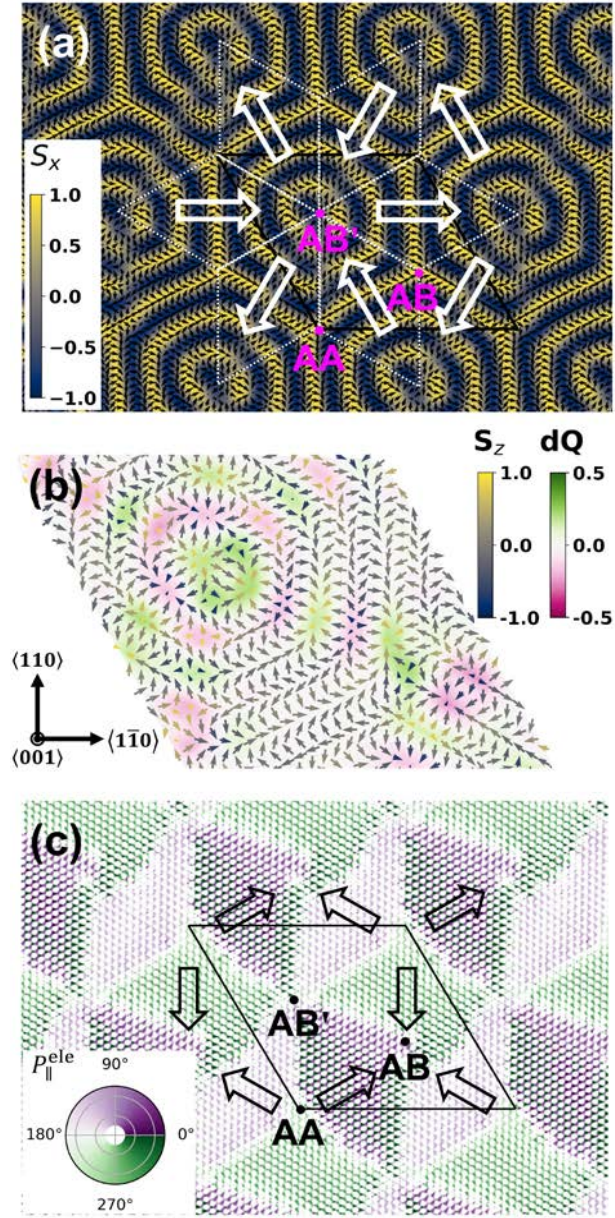

FIG. 4. Spin and polarization patterns in TBN at 2.13°. (a) Well-modulated spin pattern of top layer, obtained from spin-lattice simulations with SpinGNN++. Black arrows indicate the direction and magnitude of in-plane magnetization, with the background color mapping the $S_x$ component. Large white arrows denote the local spiral wave vectors. The white dot lines indicate the well-ordered $q$-vector patches. AFM interlayer coupling causes spin antialignment between layers, resulting in identical $q$-vectors in magnitude and direction. (b) Enlarged view of a moiré unit cell of the black parallelogram outlined in (a). The color of the arrows represents the magnitude of $S_z$, and the background color maps the local topological charge $dQ$. (c) Electric polarization pattern induced by the spin texture in (a). Small arrows denote local polarizations between nearest Ni-Ni pairs; color indicates their in-plane orientation. Large black arrows indicate net polarization directions in specific domains.

The magnetically induced electric polarization in TBN is further analyzed using the generalized KNB (GKNB) model[12], as implemented in PASP[47]. The electric dipole is induced by spin pairs via $\boldsymbol{P}_{ij} = \boldsymbol{M} \cdot \boldsymbol{S}_i \times \boldsymbol{S}_j$, with $\boldsymbol{P}$, $\boldsymbol{M}$, and $\boldsymbol{S}$ representing polarization, coupling tensor, and spin, respectively, and $i$ and $j$ denote Ni sites within the plane. The $3 \times 3$ coupling tensor $\boldsymbol{M}$ is determined via the four-state method from first-principles for spin pairs aligned along the $a$-axis. For TBN, the coupling tensor for the top (bottom) layer is obtained by rotating the monolayer tensor $\boldsymbol{M}$ by $\theta/2$ $(-\theta/2)$, given by $\boldsymbol{M}_t^{top} = \boldsymbol{R}(\theta/2)^T \cdot \boldsymbol{M} \cdot \boldsymbol{R}(\theta/2)$ and $\boldsymbol{M}_t^{bottom} = \boldsymbol{R}(-\theta/2)^T \cdot \boldsymbol{M} \cdot \boldsymbol{R}(-\theta/2)$, where $\boldsymbol{R}(\theta/2)$ and $\boldsymbol{R}(-\theta/2)$ are rotation matrices for clockwise and counterclockwise rotations, respectively. This approach is valid due to the significantly stronger intralayer interactions compared to the weaker interlayer couplings in vdW materials. The dominant components are $M_{12} = 224 \times 10^{-5} e \cdot Å$ and $M_{13} = 162 \times 10^{-5} e \cdot Å$, while other components are negligible ($<10 \times 10^{-5} e \cdot Å$). The GKNB model predicts an electric polarization in the $xy$-plane [FIG.4(c)], oriented precisely perpendicular to the in-plane spiral wave vector shown in FIG.4(a). By contrast, the standard KNB expression leads to qualitatively wrong results in certain cases [15].

Recently, electrically tunable topological magnetism, including $k\pi$-skyrmion lattices at the moiré scale, was reported in TBN[34]. In contrast, our study uncovers moiré-scale modulations in both magnetism and polarization but differs in key details. These discrepancies stem from our genuinely first-principles-accurate structural relaxations, rather than presupposing which stackings expand or contract. Moreover, we observe only topological defects at the convergence of $q$-vector patches, not skyrmions, as the actual spiral periodicity (~$5a$) is too short to stabilize an ordered skyrmion lattice without an external magnetic field.

Direct comparison reveals that the spin-driven and ionic polarizations exhibit complementary spatial distributions. In high-symmetry regions, such as AB′, strong interlayer relaxation yields a peak ionic polarization of approximately $|P_z^{ion}|_{max} \sim 10 \times 10^{-3} e \cdot Å$ [FIG.2(h)], while the local spin-driven polarization is suppressed due to interference among multiple spiral $q$-vectors. In contrast, the spin-driven component reaches $|P_\parallel^{ele}|_{max} \sim 12 \times 10^{-3} e \cdot Å$ in surrounding domains [FIG.4(c)], underscoring its slightly larger contribution to the overall moiré-scale dipolar landscape. This complementary dipole arrangement, absent in untwisted stacks, indicates that twisting unlocks a moiré-patterned magnetoelectric response that can be reversed by modest out-of-plane fields. Such emergent, field-tunable coupling offers a scalable route to electrically control complex spin textures in other twisted 2D magnets. The predicted spin textures and polarization patterns may be probed by spin-polarized scanning tunneling microscopy (SP-STM), magnetic force microscopy (MFM), and piezoresponse force microscopy (PFM). In particular, the polarization features enable real-space identification of local stacking and sliding directions in twisted bilayers.

**Polar meron-antimeron networks in anti-aligned TBN near 60°.**

For TBN near 60°, structural relaxation also leads to ionic displacements and ionic polarization (FIG. S5[45]). At $\theta = 57.87°$, a net ionic OPP of –0.079 $e \cdot Å$ develops, comparable to that of the $\theta = 2.13°$ case. In contrast, the spin–lattice molecular dynamics simulations reveal a uniform in-plane cycloidal state along the $\langle 110 \rangle$ direction with $\lambda \sim 5a$ the spin patterns, displaying no obvious changes than that of the untwisted bilayer (FIG. S11 and FIG. S12[45]).

On the other hand, the bilayer stacking ferroelectricity (BSF) theory[42] indicates polarization for bilayer $NiI_2$ at 60°. In such case, DFT results reveal that (i) the sliding-induced OPP peaks at R-AB and R-AB′ ($|P_{s,\perp}|=2.09 \times 10^{-3} e \cdot Å$), while (ii) the in-plane component $P_{s,\parallel}$ vanishes at R-AA and reaches extrema of $\pm 1.27 \times 10^{-3} e \cdot Å$ between the R-AB and R-AB′ domains [FIG. S13(a)[45]]. Similar stacking-dependent polarization behaviors have been reported in h-BN[5, 43], $MoS_2$[4, 48], and $WTe_2$[49].

Interestingly, the sliding-induced polarization form topological patterns in TBN near 60°. To extend this stacking framework to twisted systems, we map the stacking-dependent polarization field $P_s(r_0)$ onto real space using the local interlayer displacement $r_0(R) = r_0(0) + \theta \begin{bmatrix} 0 & 1 \\ -1 & 0 \end{bmatrix} R$, where $\theta$ is the twist angle and R is the real-space coordinate. While $P_{s,\perp}$ retains a triangular domain structure, $P_{s,\parallel}$ forms continuous vortex–antivortex textures around R-AB and R-AB′ regions, each with winding number $w = 1$. These textures satisfy $\nabla \cdot P_{s,\parallel}=0$ and $\nabla \times P_{s,\parallel} \neq 0$, resulting in a topological meron–antimeron network in the polarization field. Such findings indicate that stacking ferroelectricity also plays an important role in twisted systems.

In summary, we have demonstrated that the polar properties of twisted bilayer $NiI_2$ extend far beyond those of untwisted structures. Without a twist, aligned $NiI_2$ bilayers remain nonpolar because inversion symmetry is preserved during sliding. When a twist is introduced, however, structural relaxation breaks this symmetry and induces a net out-of-plane dipole, thus enabling new strategies for tuning ferroelectric order. Near $\theta \sim 2.13°$, structural and magnetic relaxations give rise to moiré-locked spin screws and in-plane polarization patches. For TBN at $\theta \sim 60°$, topological meron-antimeron networks emerge in anti-aligned bilayers, which provides a new framework for manipulating topological phases in 2D materials. Together, these findings highlight the rich interplay between spin and polarization in twisted 2D magnets, offering new opportunities to tailor emergent quantum states and laying the groundwork for applications in high-density data storage and advanced magnetoelectric devices.

**ACKNOWLEDGMENTS**

We acknowledge financial support from NSFC (No. 12188101), the National Key R&D Program of China (No. 2022YFA1402901), Shanghai Science and Technology Program (No. 23JC1400900), the Guangdong Major Project of the Basic and Applied Basic Research (Future functional materials under extreme conditions--2021B0301030005), Shanghai Pilot Program for Basic Research—FuDan University 21TQ1400100 (23TQ017), the robotic AI-Scientist platform of Chinese Academy of Science, and New Cornerstone Science Foundation.

**Reference**

[1] C. Gong, L. Li, Z. Li, H. Ji, A. Stern, Y. Xia, T. Cao, W. Bao, C. Wang, Y. Wang, Z. Q. Qiu, R. J. Cava, S. G. Louie, J. Xia and X. Zhang, Nature **546**, 265-269 (2017).
[2] B. Huang, G. Clark, E. Navarro-Moratalla, D. R. Klein, R. Cheng, K. L. Seyler, D. Zhong, E. Schmidgall, M. A. McGuire, D. H. Cobden, W. Yao, D. Xiao, P. Jarillo-Herrero and X. Xu, Nature **546**, 270-273 (2017).
[3] L. Li and M. Wu, ACS Nano **11**, 6382-6388 (2017).
[4] M. Vizner Stern, Y. Waschitz, W. Cao, I. Nevo, K. Watanabe, T. Taniguchi, E. Sela, M. Urbakh, O. Hod and M. Ben Shalom, Science **372**, 1462-1466 (2021).
[5] K. Yasuda, X. Wang, K. Watanabe, T. Taniguchi and P. Jarillo-Herrero, Science **372**, 1458-1462 (2021).
[6] F. Matsukura, Y. Tokura and H. Ohno, Nat. Nanotechnol **10**, 209-220 (2015).
[7] T. Kurumaji, S. Seki, S. Ishiwata, H. Murakawa, Y. Kaneko and Y. Tokura, Phys. Rev. B **87**, 014429 (2013).
[8] A. Pimenov, A. A. Mukhin, V. Y. Ivanov, V. D. Travkin, A. M. Balbashov and A. Loidl, Nature Phys **2**, 97-100 (2006).
[9] S. Kibayashi, Y. Takahashi, S. Seki and Y. Tokura, Nat Commun **5**, 4583 (2014).
[10] Y. Tokura, S. Seki and N. Nagaosa, Reports on Progress in Physics **77**, 076501 (2014).
[11] D. I. J. P. Khomskii, Physics **2**, 20 (2009).
[12] H. J. Xiang, E. J. Kan, Y. Zhang, M. H. Whangbo and X. G. Gong, Phys. Rev. Lett. **107**, 157202 (2011).
[13] W. Zhu, P. Wang, H. Zhu, H. Zhu, X. Li, J. Zhao, C. Xu and H. Xiang, Phys. Rev. Lett. **134**, 066801 (2025).
[14] Q. Song, C. A. Occhialini, E. Ergeçen, B. Ilyas, D. Amoroso, P. Barone, J. Kapeghian, K. Watanabe, T. Taniguchi, A. S. Botana, S. Picozzi, N. Gedik and R. Comin, Nature **602**, 601-605 (2022).
[15] W. Pan, Z. Chen, D. Wu, W. Zhu, Z. Xu, L. Li, J. Feng, B.-L. Gu, W. Duan and C. Xu, arXiv:2502.16442,  (2025).
[16] D. Amoroso, P. Barone and S. Picozzi, Nat Commun **11**, 5784 (2020).

[17] S. R. Kuindersma, J. P. Sanchez and C. Haas, Physica B+C **111**, 231-248 (1981).
[18] J. Das, M. Akram and O. Erten, Phys. Rev. B **109**, 104428 (2024).
[19] M. Amini, A. O. Fumega, H. González-Herrero, V. Vaňo, S. Kezilebieke, J. L. Lado and P. Liljeroth, Adv. Mater **36**, 2311342 (2024).
[20] X. Li, C. Xu, B. Liu, X. Li, L. Bellaiche and H. Xiang, Phys. Rev. Lett. **131**, 036701 (2023).
[21] F. Y. Gao, X. Peng, X. Cheng, E. Viñas Boström, D. S. Kim, R. K. Jain, D. Vishnu, K. Raju, R. Sankar, S.-F. Lee, M. A. Sentef, T. Kurumaji, X. Li, P. Tang, A. Rubio and E. Baldini, Nature **632**, 273-279 (2024).
[22] N. Liu, C. Wang, C. Yan, C. Xu, J. Hu, Y. Zhang and W. Ji, Phys. Rev. B **109**, 195422 (2024).
[23] Q. Liu, W. Su, Y. Gu, X. Zhang, X. Xia, L. Wang, K. Xiao, X. Cui, X. Zou, B. Xi, J.-W. Mei and J.-F. Dai, Nat Commun **16**, 4221 (2025).
[24] J. Kapeghian, D. Amoroso, C. A. Occhialini, L. G. P. Martins, Q. Song, J. S. Smith, J. J. Sanchez, J. Kong, R. Comin, P. Barone, B. Dupé, M. J. Verstraete and A. S. Botana, Phys. Rev. B **109**, 014403 (2024).
[25] P. Jiang, C. Wang, D. Chen, Z. Zhong, Z. Yuan, Z.-Y. Lu and W. Ji, Phys. Rev. B **99**, 144401 (2019).
[26] N. A.-O. Sivadas, S. A.-O. Okamoto, X. Xu, C. J. Fennie and D. Xiao, Nano Lett,  (2018).
[27] Q. Tong, F. Liu, J. Xiao and W. Yao, Nano Lett **18**, 7194-7199 (2018).
[28] M. Akram, H. LaBollita, D. Dey, J. Kapeghian, O. Erten and A. S. Botana, Nano Lett **21**, 6633-6639 (2021).
[29] M. Akram and O. Erten, Phys. Rev. B **103**, L140406 (2021).
[30] F. Xiao, K. Chen and Q. Tong, Phys. Rev. Research **3**, 013027 (2021).
[31] Y. Xu, A. Ray, Y.-T. Shao, S. Jiang, K. Lee, D. Weber, J. E. Goldberger, K. Watanabe, T. Taniguchi, D. A. Muller, K. F. Mak and J. Shan, Nat. Nanotechnol **17**, 143-147 (2022).
[32] T. Song, Q.-C. Sun, E. Anderson, C. Wang, J. Qian, T. Taniguchi, K. Watanabe, M. A. McGuire, R. Stöhr, D. Xiao, T. Cao, J. Wrachtrup and X. Xu, Science **374**, 1140-1144 (2021).
[33] H. Xie, X. Luo, Z. Ye, Z. Sun, G. Ye, S. H. Sung, H. Ge, S. Yan, Y. Fu, S. Tian, H. Lei, K. Sun, R. Hovden, R. He and L. Zhao, Nat. Phys **19**, 1150-1155 (2023).
[34] T. V. C. Antão, J. L. Lado and A. O. Fumega, Nano Lett,  (2024).
[35] K. Hejazi, Z.-X. Luo and L. Balents, PNAS **117**, 10721-10726 (2020).
[36] A. Weston, E. G. Castanon, V. Enaldiev, F. Ferreira, S. Bhattacharjee, S. Xu, H. Corte-León, Z. Wu, N. Clark, A. Summerfield, T. Hashimoto, Y. Gao, W. Wang, M. Hamer, H. Read, L. Fumagalli, A. V. Kretinin, S. J. Haigh, O. Kazakova, A. K. Geim, V. I. Fal'ko and R. Gorbachev, Nat. Nanotechnol **17**, 390-395 (2022).
[37] K. Ko, A. Yuk, R. Engelke, S. Carr, J. Kim, D. Park, H. Heo, H.-M. Kim, S.-G. Kim, H. Kim, T. Taniguchi, K. Watanabe, H. Park, E. Kaxiras, S. M. Yang, P. Kim and H. Yoo, Nat. Mater **22**, 992-998 (2023).
[38] V. Govinden, S. Prokhorenko, Q. Zhang, S. Rijal, Y. Nahas, L. Bellaiche and N. Valanoor, Nat. Mater **22**, 553-561 (2023).
[39] A. K. Yadav, C. T. Nelson, S. L. Hsu, Z. Hong, J. D. Clarkson, C. M. Schlepütz, A. R. Damodaran, P. Shafer, E. Arenholz, L. R. Dedon, D. Chen, A. Vishwanath, A. M. Minor, L. Q. Chen, J. F. Scott, L. W. Martin and R. Ramesh, Nature **530**, 198-201 (2016).
[40] S. Das, Y. L. Tang, Z. Hong, M. A. P. Gonçalves, M. R. McCarter, C. Klewe, K. X. Nguyen, F. Gómez-Ortiz, P. Shafer, E. Arenholz, V. A. Stoica, S. L. Hsu, B. Wang, C. Ophus, J. F. Liu, C. T. Nelson,

S. Saremi, B. Prasad, A. B. Mei, D. G. Schlom, J. Íñiguez, P. García-Fernández, D. A. Muller, L. Q. Chen, J. Junquera, L. W. Martin and R. Ramesh, Nature **568**, 368-372 (2019).

[41] Y. J. Wang, Y. P. Feng, Y. L. Zhu, Y. L. Tang, L. X. Yang, M. J. Zou, W. R. Geng, M. J. Han, X. W. Guo, B. Wu and X. L. Ma, Nat. Mater **19**, 881-886 (2020).

[42] J. Ji, G. Yu, C. Xu and H. J. Xiang, Phys. Rev. Lett. **130**, 146801 (2023).

[43] D. Bennett, G. Chaudhary, R.-J. Slager, E. Bousquet and P. Ghosez, Nat Commun **14**, 1629 (2023).

[44] H. Yu, B. Liu, Y. Zhong, L. Hong, J. Ji, C. Xu, X. Gong and H. Xiang, Phys. Rev. B **110**, 104427 (2024).

[45] See Supplemental Material at xxx for detailed methods and further discussions, which includes Refs. [].

[46] H. Xiang, C. Lee, H.-J. Koo, X. Gong and M.-H. Whangbo, Dalton Transactions **42**, 823-853 (2013).

[47] F. Lou, X. Y. Li, J. Y. Ji, H. Y. Yu, J. S. Feng, X. G. Gong and H. J. Xiang, J. Chem. Phys. **154**, 114103 (2021).

[48] J. Shi, P. Yu, F. Liu, P. He, R. Wang, L. Qin, J. Zhou, X. Li, J. Zhou, X. Sui, S. Zhang, Y. Zhang, Q. Zhang, T. C. Sum, X. Qiu, Z. Liu and X. Liu, Adv. Mater **29**, 1701486 (2017).

[49] Z. Fei, W. Zhao, T. A. Palomaki, B. Sun, M. K. Miller, Z. Zhao, J. Yan, X. Xu and D. H. Cobden, Nature **560**, 336-339 (2018).